\documentclass[a4paper,11pt]{article}
\usepackage{jinstpub} 
\usepackage{lineno}
\usepackage{subcaption}
\usepackage{caption}

\title{Radiation Testing of New Readout Electronics for the CMS ECAL Barrel}

\author[a]{Nico Härringer}
\author[a]{, Günther Dissertori}
\author[a]{, Tomasz Gadek}
\author[b]{, Wojciech Hajdas}
\author[a]{, Christian Haller}
\author[c]{, Nikitas Loukas}
\author[a]{, Werner Lustermann}
\author[b]{, Ljiljana Morvaj}
\author[c]{, Alexander Singovski}
\author[a]{, Krzysztof Stachon}
\author{on behalf of the CMS ECAL collaboration}
\affiliation[a]{ETH Z\"urich, Institute for Particle Physics and Astrophysics (IPA), Switzerland}
\affiliation[b]{Paul Scherrer Institute (PSI), Switzerland}
\affiliation[c]{University of Notre Dame, Department of Physics and Astronomy, United States\\}

\emailAdd{nico.harringer@cern.ch}

\abstract{In preparation of the operation of the CMS electromagnetic calorimeter (ECAL) barrel at the High Luminosity Large Hadron Collider (HL-LHC) the entire on-detector electronics will be replaced. The new readout electronics comprises 12240 very front end (VFE), 2448 front end (FE) and low voltage regulator (LVR) cards arranged into readout towers (RTs) of five VFEs, one FE and one LVR card. The results of testing one RT of final prototype cards at CERN’s CHARM mixed field facility and PSI’s proton irradiation facility are presented. They demonstrate the proper functioning of the new electronics in the expected radiation conditions.}

\notoc

\begin{document}
\maketitle
\flushbottom

\newpage

\section{\label{sec:level1}The CMS Electromagnetic Calorimeter (ECAL)}

The Compact Muon Solenoid (CMS) experiment at CERN's Large Hadron Collider (LHC) \cite{cms_experiment} features the Electromagnetic Calorimeter (ECAL) \cite{cms_ecal_tdr} to measure electron and photon energies. The ECAL barrel (EB) covers $0 < |\eta| < 1.4442$ and consists of 2448 trigger towers within 36 supermodules (SM). Each SM contains 1700 lead tungstate crystals paired with Avalanche Photodiode (APD) sensors. Groups of 25 APDs (trigger towers) connect to a motherboard with bias voltage and readout electronics, including VFE, LVR, and FE cards \cite{fe}.

\subsection{\label{sec:Upgrade}The Upgrade}

During the Long Shutdown 3 (LS3) scheduled to start in 2026, the LHC will be upgraded to the High Luminosity LHC (HL-LHC) \cite{hl_lhc_report}, which will increase its annual delivery to about 1.5 times the Run 2 dataset (138 fb$^{-1}$). Over its lifetime, it will produce roughly 10 times the LHC's integrated luminosity (4000 fb$^{-1}$). This upgrade will result in higher radiation levels and event rates, with pileup rising to 140-200 interactions per bunch-crossing (up from 40-60) \cite{cms_phase2_tdr}, and the Level-1 trigger rate increasing to 750 kHz (from 100 kHz). Enhanced granularity and timing resolution will be necessary to identify primary vertices amid this higher pileup. Additionally, radiation-hard on-detector electronics will be required. In the upgraded system, the APD signal is amplified by the Calorimeter transimpedance amplifier (CATIA, \cite{catia}) chip on the VFE cards. This pulse is digitized by the Lisbon-Torino ECAL Data Transmission Unit (LiTE-DTU, \cite{litedtu}), with data from each chip collected by four low power Gigabit transmission (lpGBT, \cite{lpgbt_manual}) chips on the FE card and transmitted via optical links with 10.26 Gbps to the off-detector Barrel Calorimeter Processor (BCP, \cite{bcp_v1}) for digital processing and trigger generation.

\section{\label{sec:testsetup_irrad}Irradiation facilities}

Positioned in front of the Hadronic Calorimeter (HCAL) \cite{cms_hcal_tdr}, ECAL will be exposed to high-energy hadrons (HEH) like protons, neutrons, and pions. By the end of its lifetime, the most affected readout tower in a SM, located at $|\eta|$ = 1.431, is expected to experience a HEH fluence of $4.5 \times 10^{13}\ \mathrm{cm}^{-2}$ and a total ionizing dose (TID) of about 5.70 kGy (7.65 kGy) corresponding to 3000 fb$^{-1}$ (4000 fb$^{-1}$) \cite{BRIL_Note}. All HL-LHC key figures discussed in this section refer to this specific position. For this campaign, two irradiation facilities were chosen each probing different aspects of the readout system.

\subsection{\label{sec:testsetup_charm}CHARM}

The ideal facility for the test offers a uniform radiation field with a sufficient dose rate and particle fluence that irradiates every part of the tower with the same rate. For these reasons, the CERN High Energy Accelerator Mixed Field (CHARM) facility \cite{charm_simulation} was chosen. Mixed field facilities like CHARM offer diverse particle species and can simulate the radiation environment in terms of energy spectra and particle mixture of CMS. A position was chosen that resulted on average in 41.3\% neutrons, 19.5\% protons, 35.6\% pions, 3.41\% kaons and 0.19\% other particles. The total time spent at CHARM including the access times was 17 days. This allowed for extensive data taking.

\subsubsection{\label{sec:data_taking_charm}Data taking and results} 

During the test, the readout tower remained powered and operational most of the time. The readout routine conducts a comprehensive system test of the new EB electronics and simulating HL-LHC operation. Critical measurements include the evolution of dark current and Single Event Upset (SEU) corrections as well as performing stability studies on baselines (pedestals) and CATIA test pulse amplitudes, while secondary measurements encompass the CATIA RMS noise evolution. The main aim of the test is to verify that the mentioned critical parameters are stable with respect to increasing radiation and that the tower experiences no critical failure in data readout or stability until the end. For a dark current readout with the GBT-SCA chip, the APDs are biased with +400V. Every six hours the tower was power cycled and every hour a reconfiguration was performed, which corresponds in total to the expected number of power cycles per year (minimum three). Every twelve seconds pedestals, CATIA test pulses, the dark current of the APDs and the SEU counter of both CATIA and lpGBTs is recorded. The CATIA test pulse simulate signals that occur from interesting physics events. The LiTE-DTU's gain switch feature automatically adjusts the gain when the signal exceeds a predefined threshold. To test this, CATIA pulses are taken for both gains. A 12-bit register controlled current injection, calibrated for high and low gain settings.

\begin{figure}[h!]
\centering
\begin{subfigure}{0.49\textwidth}
    \centering
    \includegraphics[height=4.7cm]{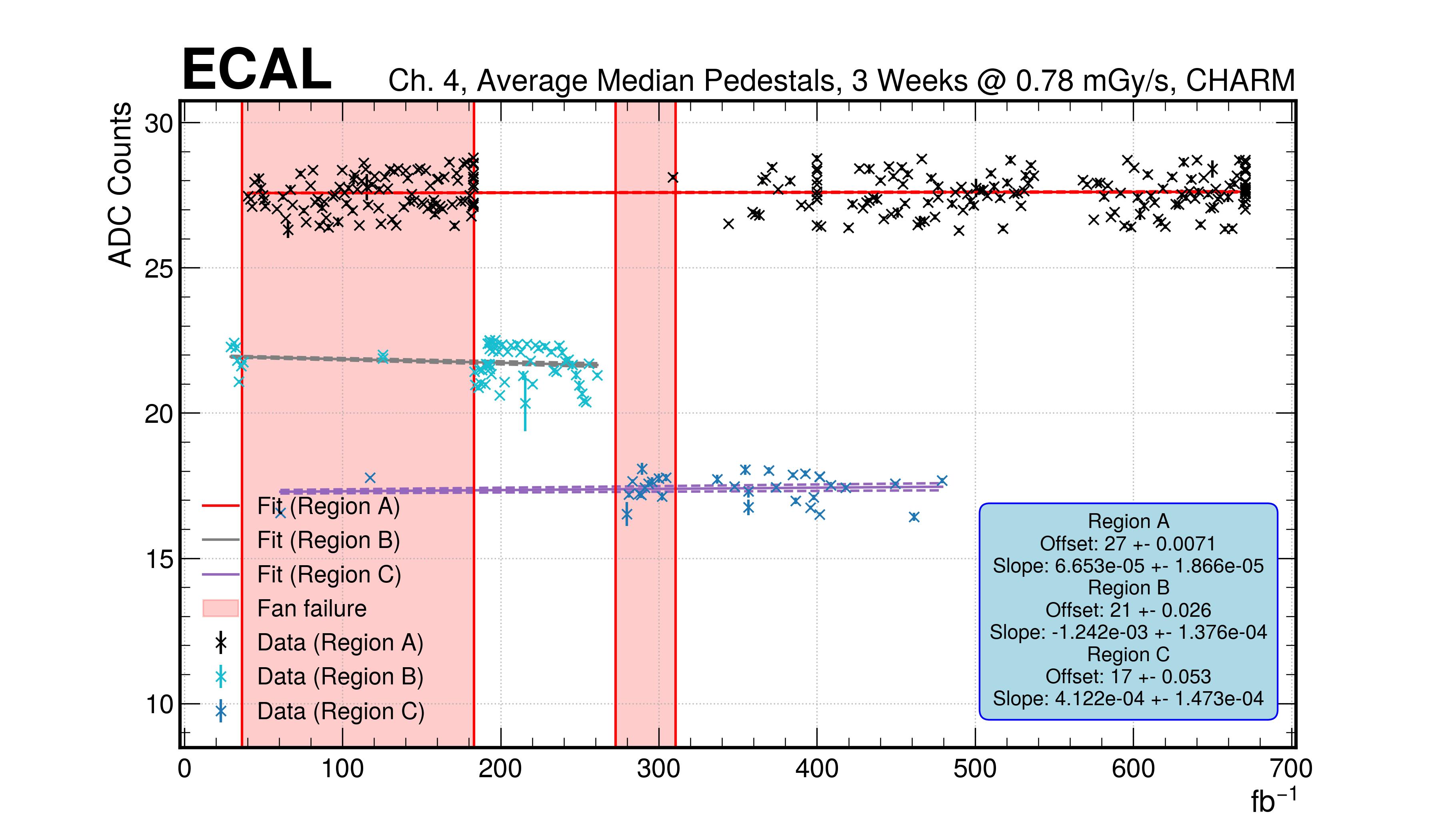}
    \caption{\label{fig:ped_4} The pedestal distribution of channel 4 as a function of the equivalent integrated luminosity. The data set is classed in three regions A, B and C corresponding each to a specific baseline subtraction value chosen during calibration.}
\end{subfigure}
\hfill
\begin{subfigure}{0.49\textwidth}
    \centering
    \includegraphics[height=4.7cm]{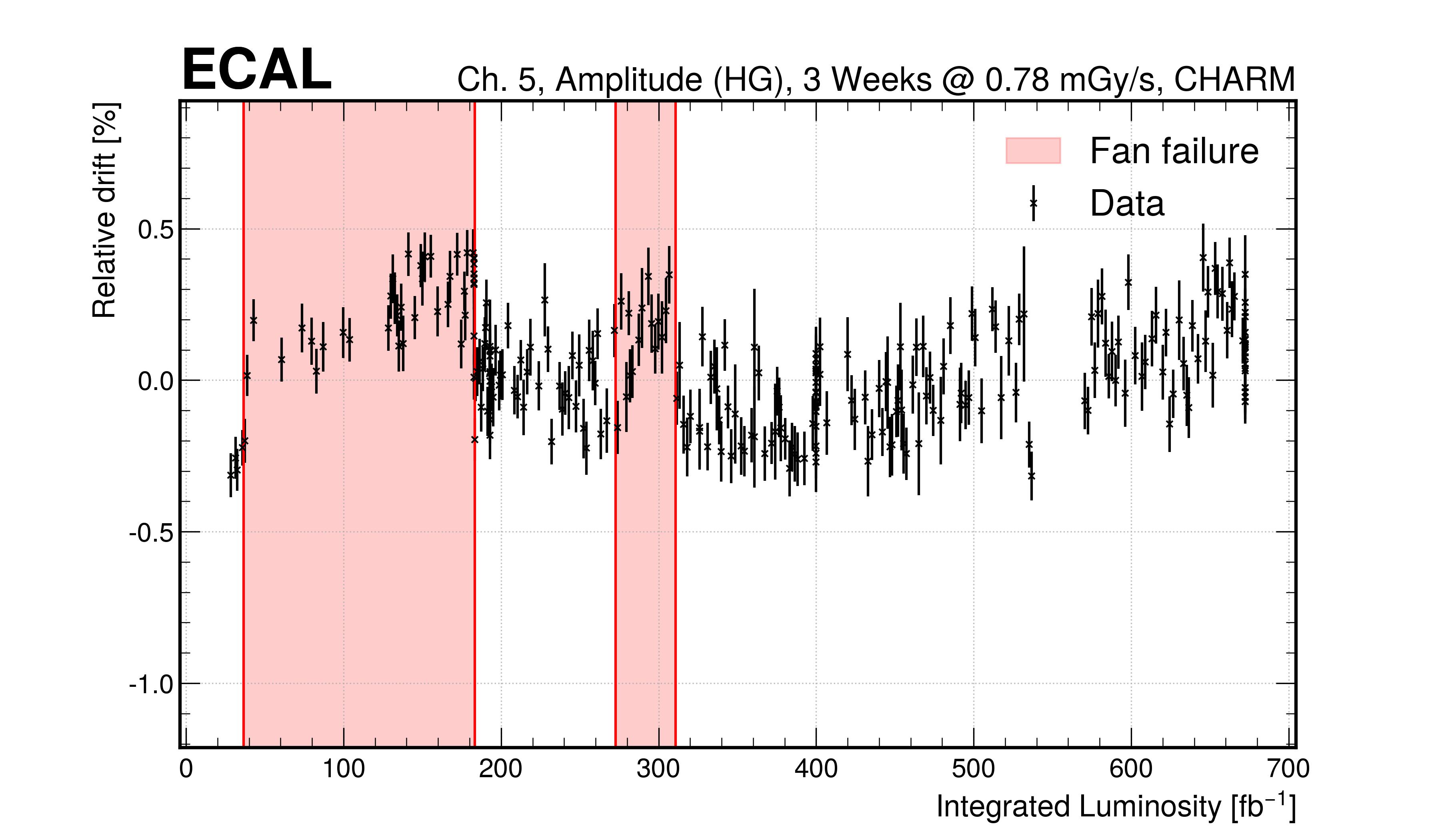}
    \caption{\label{fig:tp_hg_average_5} The relative drift of the test pulse amplitude in high gain of channel 5 test as a function of the equivalent integrated luminosity. The used reference voltage was 1.355V. The data set was not categorized since the ADC offset is small compared to the signal height.}
\end{subfigure}
\caption{The pedestal and test pulse amplitude distributions of channel 5 and 7 respectively with data from CHARM. The fan downtime window is marked in red. A factor of 0.486 is used to convert from TID to the equivalent integrated luminosity.}
\label{fig:charm_ped_tp}
\end{figure}

Every reconfiguration cycle the VFEs were calibrated. During this step, the CATIA digital-to-analog (DAC) converter register (6-bit) and the baseline subtraction register (12-bit) of the LiTE-DTU were adjusted so that the pedestal values are on average 30 $\pm$ 10 ADC counts, well below the LiTE-DTUs signal threshold of 64 counts. The exact value varies with each tower reconfiguration. The pedestals and CATIA test pulses were disentangled from the baseline set during VFE calibration and categorized based on their baseline subtraction value. During the first week at CHARM, both cooling fans malfunctioned and were replaced in the following maintenance window. The second downtime window was shorter as the fans were successfully restarted. Data collected equivalent to 673 fb$^{-1}$ suggests that pedestal values for all channels are mostly stable. The average median distributions remained constant with respect to integrated luminosity for a given CATIA / LiTE-DTU configuration as illustrated for channel 4 in Fig. \ref{fig:ped_4}. Test pulse amplitude stability was also assessed. Test pulses in high and low gain were alternated. The TID and the HEH fluence are measured with the radiation monitoring device of the facility. A factor of 0.486 is used to convert the TID to the equivalent integrated luminosity accounting for simulation uncertainties with a safety margin. A drift in amplitude indicate possible radiation effects on the channel. Channels with CATIA v2.1 performed well, showing no significant drift with relative deviations from the average test pulse amplitude under 1\% for both gains (see Fig. \ref{fig:tp_hg_average_5}).

\begin{figure}[h!]
\centering
\begin{subfigure}{0.49\textwidth}
    \centering
    \includegraphics[height=5cm]{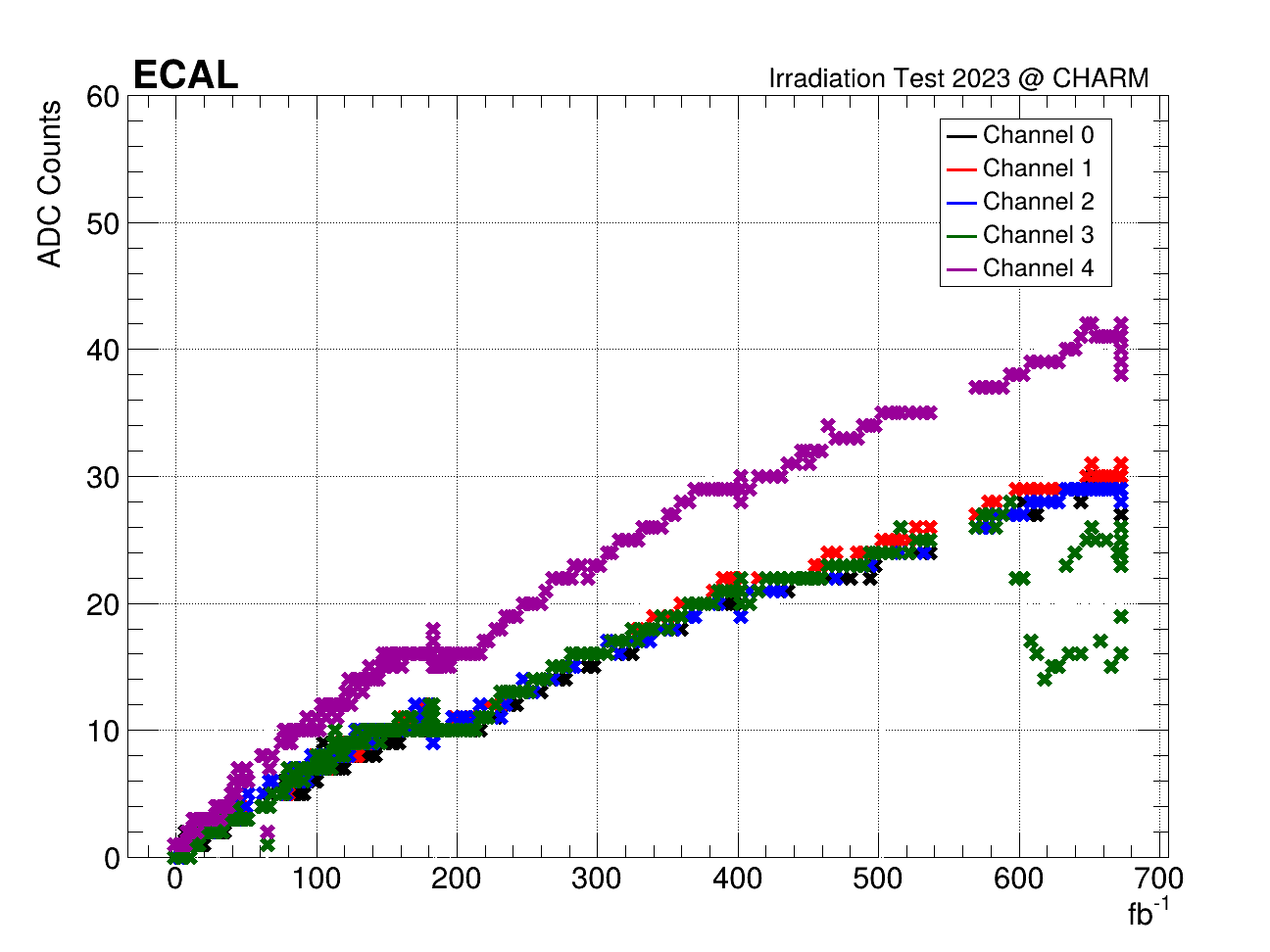}
    \caption{\label{fig:dk_full} The dark current of channels 0-4 as a function of the equivalent integrated luminosity. Each channel carries a previously not irradiated APD capsule, constantly biased with +400V.}
\end{subfigure}
\hfill
\raisebox{0.9cm}{ 
\begin{subfigure}{0.49\textwidth}
    \centering
    \includegraphics[height=4.72cm]{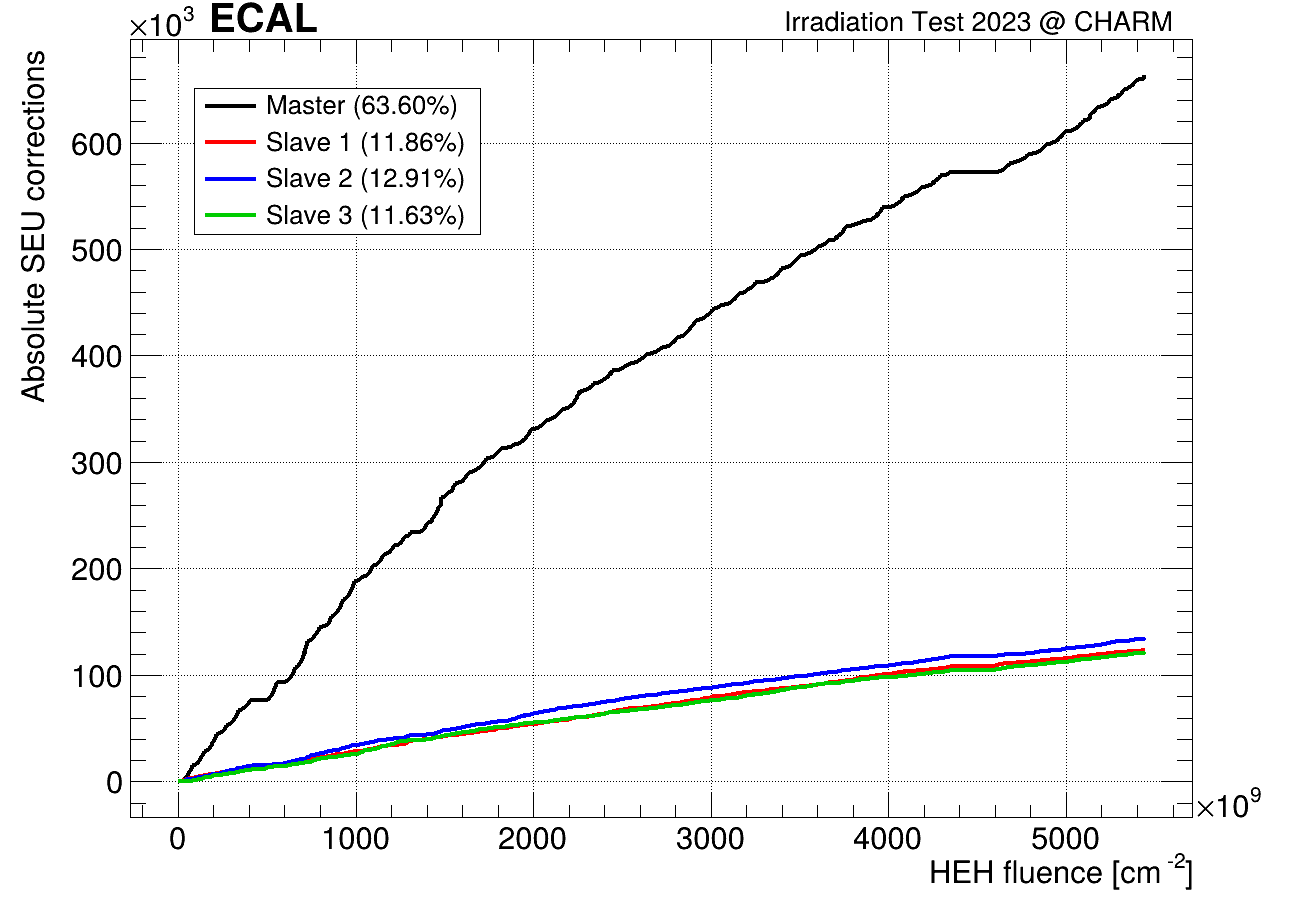}
    \caption{\label{fig:lpgbt_seu} Absolute SEU corrections as function of the HEH fluence in the CHARM test.}
\end{subfigure}
}
\caption{Comparison of dark current and SEU corrections. A factor of 0.486 is used to convert from TID to the equivalent integrated luminosity.}
\label{fig:dk_seu}
\end{figure}

To determine the raw Root Mean Square (RMS) noise, the median value $\mathrm{ped}_i$ was calculated for each channel in a pedestal dataset $i$. By subtracting $\mathrm{ped}_i$ from every data point $ADC_{i}^{j}$ in the pedestal dataset $i$, the distribution $n_{i}^{j} := ADC_{i}^{j}\ $–$\ \mathrm{ped}_i$ was obtained. The standard deviation of $n_{i}^{j}$ is referred to as the raw RMS noise, denoted as $\sigma_{\mathrm{rawnoise}, i}$. The objective of this study was to verify the stability of $\sigma_{\mathrm{rawnoise}, i}$ with respect to radiation. High raw RMS noise degrades data quality and is undesired. The evolution of the raw RMS noise in relation to the TID for all 25 channels was calculated. When APDs are connected, the raw RMS noise adds to the increasing dark current of the APDs. Therefore, channels affected by APD dark current were excluded from the analysis. The mean value of the RMS noise standard deviation distribution is 5.24 $\pm$ 0.92\%, while the mean of the RMS noise mean distribution is 1.17 $\pm$ 2.84\%. This indicates that channels equipped with CATIA v2.1 exhibit excellent resilience to radiation up to 673 fb$^{-1}$ in terms of maintaining consistent RMS noise levels. 

The GBT-SCA chip on the FE card reliably read the dark current at all times during the test. Figure \ref{fig:dk_full} shows the dark current's evolution with equivalent integrated luminosity. The APD capsules in channels 0-3 were uniformly irradiated (see Fig. \ref{fig:dk_full}). However, channel 4 displayed a different behavior, likely due to its capsule being positioned unfavorably within the beam box, leading to uneven radiation exposure. Alternatively, the capsule could be inherently more susceptible to radiation. APD 3 experienced contact issues near the test's end, causing the readings to oscillate.

\subsubsection{\label{sec:results_SEU_charm}Single Event Upsets (SEU)}

A Single Event Upset (SEU) is a transient error in a digital circuit caused by a high-energy particle interacting with a semiconductor. All chips used in the tower use register triplication and a majority voter to mitigate the effects of SEUs on the operation. The objective was to confirm that SEU mitigation prevents operational disruptions and ensures data integrity for all four lpGBTs \cite{lpgbt_seminar}. Failure of the mitigation could result in corrupted data, impacting the accuracy of the recorded physics. The majority of SEU corrections (63.6\%) occurred in the master lpGBT due to its role as the communication hub for slave lpGBTs \cite{lpgbt_manual, fe}. CATIA SEU  availability was restricted to channels with CATIA v2.1 due to differences in SEU logging capabilities across versions. Figure \ref{fig:lpgbt_seu} shows the absolute SEU corrections for the four lpGBT. No freezing of the lpGBT or the CATIAs was observed and they were operational at all times suggesting, that the SEU mitigation is working as intended. 

\begin{figure}[h!]
\centering
\begin{subfigure}{0.49\textwidth}
    \centering
    \includegraphics[height=4.7cm]{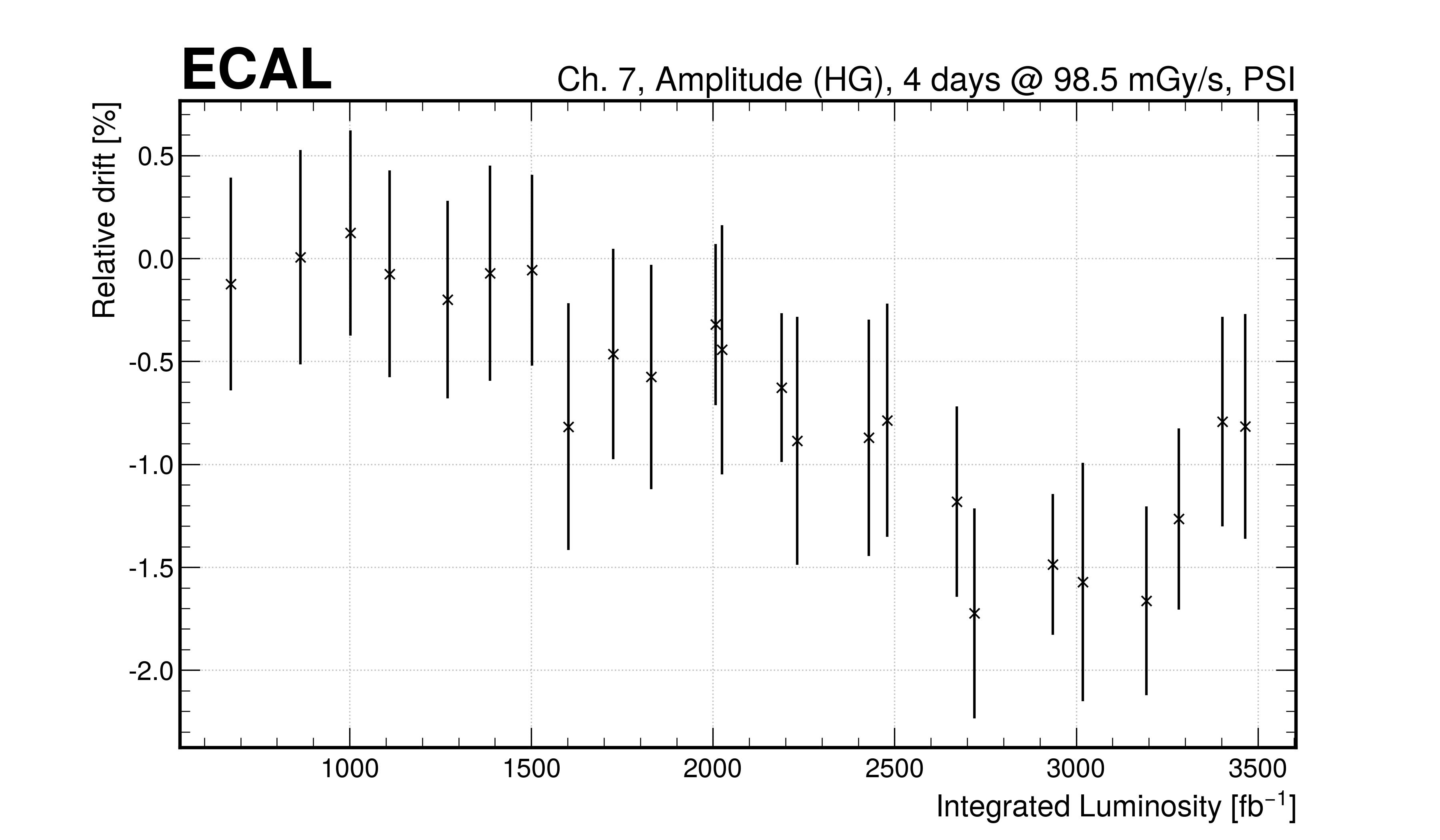}
    \caption{\label{fig:tp_hg_7_psi} The relative drift of the test pulse amplitude of channel 7 during the PSI test as a function of the equivalent integrated luminosity. The used reference voltage was 1.355V. The data set was not categorized. A factor of 0.486 is used to convert from TID to the equivalent integrated luminosity.}
\end{subfigure}
\hfill
\raisebox{0.45cm}{ 
\begin{subfigure}{0.49\textwidth}
    \centering
    \includegraphics[height=4.7cm]{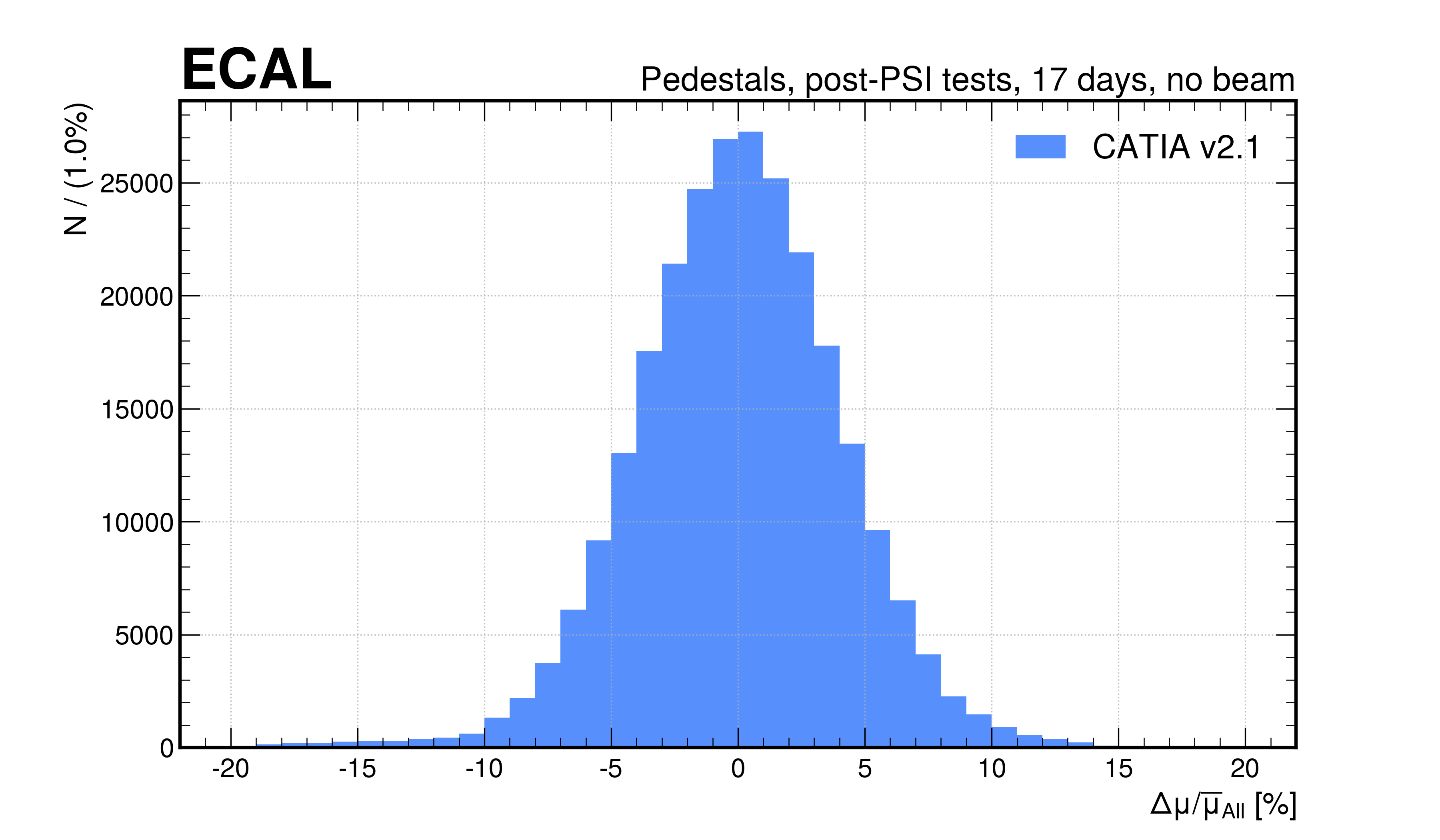}
    \caption{\label{fig:relMean_all_channels_ped} The relative deviation of the mean values of the pedestals of all CATIA v2.1 channels after PSI from the mean of the complete dataset. In total 260945 datapoints were collected. The mean of the histogram is 0 $\pm$ 4.07\%.}
\end{subfigure}
}
\caption{Comparison of test pulse amplitude drift and pedestal mean deviation.}
\label{fig:side_by_side_psi}
\end{figure}

\subsection{\label{sec:testsetup_psi}PSI}

To achieve the target dose of 5.7 kGy (3000 fb$^{-1}$) and potentially 7.65 kGy (4000 fb$^{-1}$), a more intense particle source is required allowing also to stress test the system beyond its designed limits. Low energy photons like those from Co-60 sources are negligible, as ECAL itself shields from the direct electromagnetic load. The major radiation on ECAL comes from the tails of hadron showers, including reflected hadron showers from HCAL. The Proton Irradiation Facility (PIF) at the Paul Scherrer Institute (PSI) in Switzerland provides directed high-energy protons from the COMET cyclotron \cite{psi_pif} and it was also used for testing the legacy ECAL trigger towers \cite{legacy_psi}. 74 MeV protons were used, yielding a proton flux of $8.70 \times 10^7 \ \mathrm{cm}^{-2} \mathrm{s}^{-1}$. With a stopping power of $\frac{dE}{dx} = 7.326\ \mathrm{\frac{MeV \ cm^2}{g}}$ in silicon, the expected TID per hour is about 356 Gy, with a proton fluence of $3.13 \times 10^{11} \ \mathrm{cm}^{-2}$ and a 1 MeV equivalent neutron fluence of $3.19 \times 10^{11} \ \mathrm{cm}^{-2}$. The beam profile of COMET is Gaussian, leading to a non-uniform irradiation of the tower. Over four night shifts (16.19 hours), a TID of 5.77 kGy was accumulated, with an average dose rate of 99.0 mGy/s for the front center of the tower. As in CHARM, a factor of 0.486 is applied to convert the TID to the equivalent integrated luminosity. The newest version of the FE card was used (v3.3). The VFE cards were oriented tangentially to the beam. This exposure resulted in dose rates varying across channels, with values ranging from 61.4 mGy/s (channel 20) to 98.5 mGy/s (channels 7 and 12).

\subsubsection{\label{sec:data_taking_psi}Data taking and results}

The data taking routine followed the CHARM approach, including periodic tower reconfiguration and monitoring of critical parameters, but with modifications to save time. Due to the limited time available, less data was collected. The power cycling feature was removed. Every 30 minutes the tower was reconfigured. Combined with the dose from CHARM, a cumulated TID of 7.15 kGy and a proton fluence of $4.92 \times 10^{12} \ \mathrm{cm}^{-2}$ was achieved. Similarly, the proton fluence target was reached by 83.40\%, and the $4.42 \times 10^{12} \ \mathrm{cm}^{-2}$ target for 3000 fb$^{-1}$ by 111.3\%. The tower collected a high-energy hadron fluence of $5.44 \times 10^{12} \ \mathrm{cm}^{-2}$ from the CHARM test. Due to tighter time constraints at PSI, fewer data points were collected. Channels 7 and 12 received the highest radiation exposure, approximately 98.5 mGy/s. Test pulse stability was assessed by comparing amplitudes across various gain settings, following the same procedure as in CHARM. For CATIA v2.1, the relative drift of the test pulse amplitude remained within 2\% in both gains. Specifically, channel 7 showed a relative drift of 1.7\% in high gain (see Fig. \ref{fig:tp_hg_7_psi}) and 1.5\% in low gain, while channel 12 had a drift of 5\% in high gain and 4\% in low gain. The mean value of the RMS noise distribution is 0.52 $\pm$ 2.66\%, while the mean of the RMS noise standard deviation distribution is 4.54 $\pm$ 1.13\%. No malfunction of the lpGBTs and the CATIAs occured. 

\section{\label{sec:discussion}Discussion}

Both the CHARM and PSI irradiation campaigns confirmed that the on-detector electronics for the CMS ECAL upgrade for HL-LHC can withstand extreme conditions. The PSI RMS results are in agreement with CHARM within 1$\sigma$ indicating resilience against high proton fluxes and dose rates well beyond the designed limits for HL-LHC. The test pulse amplitude drift at PSI is on average 1\% higher than in CHARM, very likely originating from the higher dose rate. The communication between the readout tower and backend was flawless, with a bit error rate of $2.87 \times 10^{-14}$ over a 1-hour test across all four lpGBTs. Throughout both tests, the GBT-SCA chip consistently read out the dark current without any SEU issues, ensuring continuous channel monitoring. Four months post-irradiation, a 17-day re-evaluation at CERN (without beam) over 17 days showed no channel drift, indicating long-term stability (Fig.  \ref{fig:relMean_all_channels_ped}). CATIA v2.1 demonstrated high precision in handling injected test pulses, even under dose rates up to 98.5 mGy/s. These results suggest that the upgraded readout electronics will perform reliably throughout the HL-LHC's operational period.

\begin{acknowledgments}
The authors would like to thank Giacomo Cucciati, Tanja Gisler and the ECAL community for their valuable help and expertise in making these irradiation test campaigns work. A big thank you also the PSI beam operators, as well as to the staff from CHARM.
\end{acknowledgments}




\bibliographystyle{JHEP}
\bibliography{bibliography.bib}
\end{document}